\documentclass[conference]{IEEEtran}
\IEEEoverridecommandlockouts
\usepackage{cite}
 \pdfoutput=1 
\usepackage{listings}
\usepackage{amsmath,amssymb,amsfonts}
\usepackage{algorithmic}
\usepackage{graphicx}
\usepackage{textcomp}
\usepackage{multirow}
\usepackage{xcolor}
\usepackage{diagbox}

\usepackage{subcaption}
\def\BibTeX{{\rm B\kern-.05em{\sc i\kern-.025em b}\kern-.08em
    T\kern-.1667em\lower.7ex\hbox{E}\kern-.125emX}}
\begin{document}

\title{Side Channel-Assisted Inference Leakage from Machine Learning-based ECG Classification}

\author{\IEEEauthorblockN{Jialin Liu\IEEEauthorrefmark{1},
Han Wang\IEEEauthorrefmark{1},
Najmeh Nazari\IEEEauthorrefmark{1}, 
Behnam Omidi\IEEEauthorrefmark{2},
Avesta Sasan\IEEEauthorrefmark{1},\\
Khaled N. Khasawneh\IEEEauthorrefmark{2},
Setareh Rafatirad\IEEEauthorrefmark{1},and 
Houman Homayoun\IEEEauthorrefmark{1}}
\IEEEauthorblockA{\IEEEauthorrefmark{1}
University of California, Davis\\
Email: \{czfang,hjlwang,nnazaribavarsad,asasan,srafatirad,hhomayoun\}@ucdavis.edu}
\IEEEauthorblockA{\IEEEauthorrefmark{2}
George Mason University\\
Email: \{bomidi,kkhasawn\}@gmu.edu}
}

\author{\IEEEauthorblockN{Jialin Liu}
\IEEEauthorblockA{\textit{Electrical and Computer Engineering} \\
\textit{Temple University}\\
jialinliu@temple.edu}
\and
\IEEEauthorblockN{Ning Miao}
\IEEEauthorblockA{\textit{Electrical and Computer Engineering} \\
\textit{University of California, Davis}\\
nmiao@ucdavis.edu}
\and
\IEEEauthorblockN{Chongzhou Fang}
\IEEEauthorblockA{\textit{Electrical and Computer Engineering} \\
\textit{University of California, Davis}\\
czfang@ucdavis.edu}
\and
\IEEEauthorblockN{Houman Homayoun}
\IEEEauthorblockA{\textit{Electrical and Computer Engineering} \\
\textit{University of California, Davis}\\
hhomayoun@ucdavis.edu}
\and
\IEEEauthorblockN{Han Wang}
\IEEEauthorblockA{\textit{Electrical and Computer Engineering} \\
\textit{Temple University}\\
han.wang.hw@temple.edu}
}

\maketitle

\begin{abstract}
The Electrocardiogram (ECG) measures the electrical cardiac activity generated by the heart to detect abnormal heartbeat and heart attack. However, the irregular occurrence of the abnormalities demands continuous monitoring of heartbeats. Machine learning techniques are leveraged to automate the task to reduce labor work needed during monitoring. In recent years, many companies have launched products with ECG monitoring and irregular heartbeat alert. Among all classification algorithms, the time series-based algorithm dynamic time warping (DTW) is widely adopted to undertake the ECG classification task. Though progress has been achieved, the DTW-based ECG classification also brings a new attacking vector of leaking the patients' diagnosis results. This paper shows that the ECG input samples' labels can be stolen via a side-channel attack, Flush+Reload. In particular, we first identify the vulnerability of DTW for ECG classification, i.e., the correlation between warping path choice and prediction results. Then we implement an attack that leverages Flush+Reload to monitor the warping path selection with known ECG data and then build a predictor for constructing the relation between warping path selection and labels of input ECG samples. Based on experiments, we find that the Flush+Reload-based inference leakage can achieve an 84.0\% attacking success rate to identify the labels of the two samples in DTW.

\end{abstract}

\section{Introduction}
\label{sec:intro}
Electrocardiogram (ECG) is the heart's electrical activity and is extensively utilized in detecting and diagnosing cardiac arrhythmias. Traditionally, well-trained cardiologists inspect the ECG waveform visually to detect abnormalities. However, the intermittent occurrences of arrhythmia impede continuous detection and diagnosis. To address the challenge, machine learning-based classification algorithms have been extensively adopted to detect an abnormal heartbeat in ECG traces \cite{huang2002ecg,tuzcu2005dynamic}. With the advancement of machine learning, many companies have launched products with ECG monitoring and irregular heart rhythm detection. Such functionality significantly benefits people in keeping track of their health condition and provides more insights for doctors to confirm a diagnosis. However, machine learning-based ECG classification also brings new security challenges. The computations in machine learning algorithms depend on the labels of ECG traces, which can be maliciously used to infer samples' labels.

\begin{figure}
    \centering
    \includegraphics[width=0.38\textwidth]{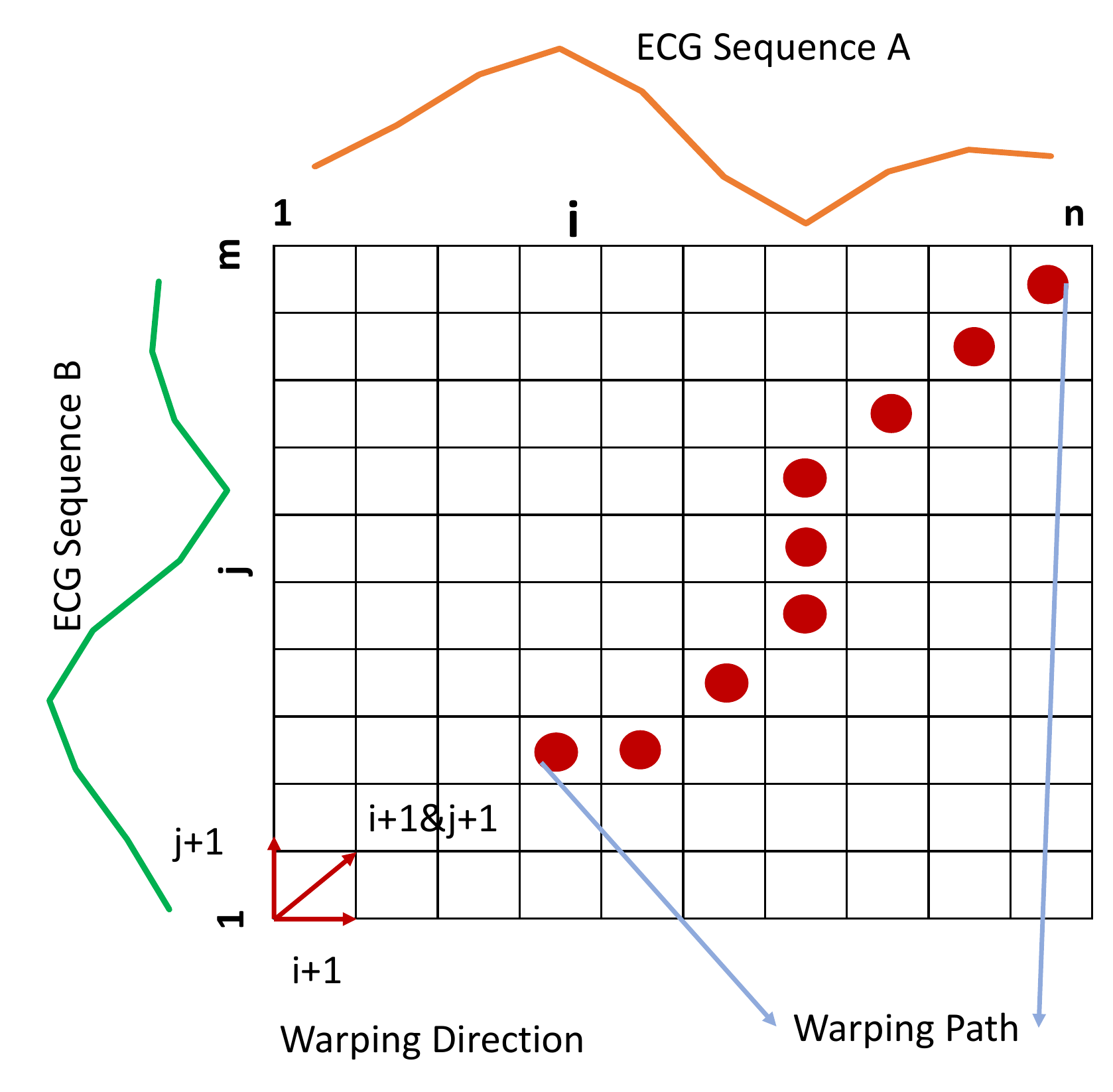}
    \caption{Warping directions and paths of DTW.}
    \label{fig:dtw}
\end{figure}

In this work, we choose one of the most prevalent machine learning algorithms, i.e., the time-series classification algorithm dynamic time warping (DTW) \cite{berndt1994using}, as a case study to demonstrate the severe privacy breach threat. DTW has long been used to identify the normality of ECG traces \cite{zhang2009electrocardiogram} due to its robustness even with scaled and shifted datasets. DTW first finds an optimal alignment of two temporal sequences by allowing a nonlinear mapping of one sequence to another to minimize the distance between two sequences. Then it calculates the Euclidean distance with the aligned points in two traces. Taking two ECG temporal sequences as an example shown in Figure \ref{fig:dtw}, the two ECG sequences are presented as $A$  ($a_1$, $a_2$,..., $a_n$) and $B$ ($b_1$, $b_2$,..., $b_m$). Data points in sequence A $i$ and sequence B $j$ have three warping choices: $i+1$, $j+1$, and $i+1\&j+1$, which are called \textbf{warping direction}. The selection among the three warping directions is chosen based on Eq. \ref{eq:dtwformulation}, where the warping direction giving minimal distance will be selected. All warping directions in one round DTW calculation for sequence A and sequence B are called \textbf{warping path} or \textbf{alignment path}. Afterward, the total distance between A and B will be calculated based on the warping path and can be used for classification.

\begin{equation}
 \gamma(i,j)=(a_i, b_j)^2+\min\left\{
 \begin{array}{c}
      \gamma(i-1,j)  \\
     \gamma(i-1,j-1)\\
     \gamma(i,j-1)\\
 \end{array}
 \right.
 \label{eq:dtwformulation}
\end{equation}

We find that the temporal ECG sequences from the same class sent to the DTW model are likely to have more similar warping paths. We find that the computations of choosing among $\gamma(i-1, j)$, $\gamma(i-1, j-1)$ and $\gamma(i, j-1)$ in Eq. (\ref{eq:dtwformulation}) are dependent on the labels of ECG traces. The phenomenon is because sequences from the same labels have a high chance of similar warping selections. Though the computation behaviors in DTW do not disclose any private information directly, the correlation between the computation behaviors, i.e., warping paths and labels, presents a new vulnerability of leaking inputs' labels to attackers by spying on the warping alignment.

Based on the correlation between warping paths and labels of ECG samples, we further leverage side-channel attack (SCA) Flush+Reload to monitor the warping directions and deduce the labels of inputs sent to DTW. However, the three warping directions in most DTW implementations have close memory locations, while Flush+Reload cannot monitor all of them. To address the challenge, we propose to monitor one warping direction only and explore different types of machine learning classifiers to demonstrate the attacker's full potential for recovering labels. We utilize ECG datasets from the UC Riverside Time Series repository \cite{UCRArchive2018} and find that Flush+Reload-based attack on DTW-based ECG classification can successfully deduce inputs' labels up to 84.0\%. 

The main contributions of this work are summarized below.

\begin{itemize}
    \item To the best of our knowledge, this is the first work to explore the side channel-assisted inference leakage of DTW-based ECG classification and present the correlation between warping paths and the two input sequences' labels.
    \item In this paper, we further prototype a Flush+Reload-based inference attack that can monitor the warping directions and deduce labels of inputs.
    \item We also evaluate the importance of the three warping directions in terms of differentiating the labels, which gives insights into future defense approaches.
   
\end{itemize}

\begin{figure}[!thb]
    \centering
    \includegraphics[width=0.43\textwidth]{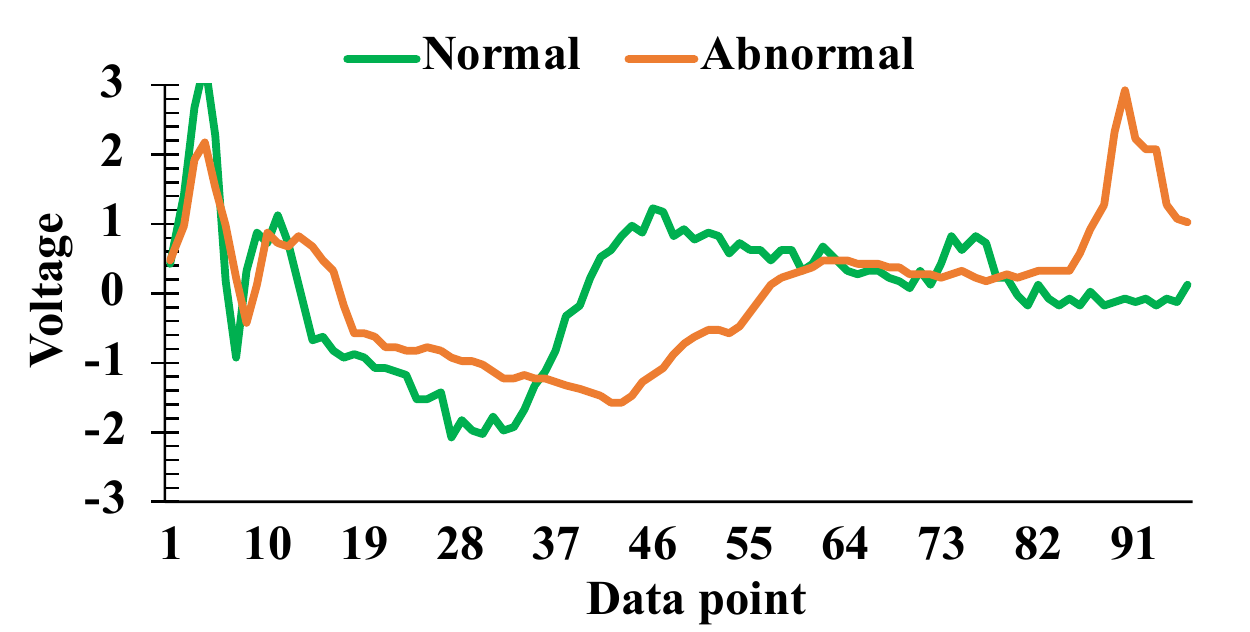}
    \caption{ECG normal and abnormal samples from ECG200 \cite{olszewski2001generalized} dataset.}
    \label{fig:ecgsample}
\end{figure}

\section{Background and Related Work}
\label{sec:bg}
\subsection{Electrocardiogram (ECG) }
An electrocardiogram (ECG) is a standard test for finding abnormal heart rhythms or cardiac (heart) diseases. Traditionally, medical practitioners identify abnormalities visually in patients' hardcopy recordings, which notably obstruct long-term monitoring for people at higher risks of heart diseases, like smoking. However, the high mortality rate of heart diseases demands continuous ECG monitoring for early detection and precise classification. Both industry and academia have studied machine learning-based ECG classification to address the issue. Apple, Samsung, and Huawei have launched ECG monitoring and classification products in recent years. It provides substantial guidance for doctors and makes tracking easier. Figure \ref{fig:ecgsample} presents one normal sample, and one abnormal ECG sample from ECG 200 dataset \cite{olszewski2001generalized}. The Y-axis presents the voltage strength of ECG signals, and X-axis presents the data points. The length of ECG samples is 100.

\subsection{Cache-based Side-Channel Attacks}
Cache hierarchies are introduced to bridge the latency gap between memory and CPU, which are shared among applications and users. The shared cache component allows attackers to observe victim applications' behaviors by causing contentions and measuring accessing time to infer victims' data access patterns. Different types of cache-based SCAs are developed in the past decade, including Flush+Reload \cite{yarom2014flush}, Flush+Flush \cite{gruss2016flush+}, Prime+Probe \cite{liu2015last}, which spy on shared cache activities and steal secrets like passwords and secret keys. In this work, we leverage \emph{Flush+Reload} to observe the computation behaviors of the victim DTW ECG classification model. \emph{Flush+Reload} \cite{yarom2014flush} exploits the vulnerability of the page de-duplication technique by monitoring the memory access lines in the shared pages. This attack targets the \emph{Last-Level Cache} in the CPU and flushes out victim applications' data in the cache continuously. Then it tries to access the data and measures the accessing time (latency). The length of the accessing time determines whether victim applications have accessed the data.
\subsection{Related Work}
Machine learning has received noteworthy progress in the past decade and has been adopted in various critical domains, such as disease diagnosis \cite{ribeiro2020automatic,hassantabar2020diagnosis}, intelligent surveillance \cite{zhou2018cyber,jin2015real}, financial decision \cite{jeong2019improving}, and so on. The security and privacy issues have raised increasing attention. Prior works have shown that leveraging hardware-oriented attacks can recover architectures, parameters, and inputs \cite{hong2018security,hua2018reverse,yan2020cache,luo2020stealthy,wei2018know,liu2021methodology,patwari2022dnn}.
\cite{liu2021methodology,wang2022stealthy}. For example, Hong et al. \cite{hong2018security,hong20200wn} leverage cache-based side-channel attack, Flush+Reload, and monitor the layers used in the inference phase. It successfully recovered MalConv and ProxylessNAS-CPU with 0\% error. Hua et al. \cite{hua2018reverse} investigates the leakage from the memory side channel on hardware accelerators and finds the memory access patterns can reverse-engineer both structures and weights of CNN models. Besides the machine learning designs, the inputs and labels are also threatened. Xiang et al. \cite{wei2018know} utilizes power side channel to reconstruct the input images on FPGA-based accelerators. The capability of the attack is evaluated in one of the most prominent image datasets, MNIST dataset \cite{lecun1998mnist}, achieving 89\% accuracy. Liu et.al \cite{liu2021methodology} presented that using software-based side channel telemetries can help infer the output class of input images. Similar work has been done by Luo et al. \cite{luo2020stealthy}, which builds a correlation between cache access patterns from Prime+Probe and inference results with statistical learning models. The deduced inference results can further help to reveal a vehicle's route or location with the adaptive Monte-Carlo localization (AMCL) algorithm. 

The above works have presented the leakage of machine learning in the face of side-channel attacks. While they lack the analysis of victim models' internal behaviors and the incurred vulnerability. In comparison, our work aims to provide vulnerability analysis and a proof-of-concept attack with Flush+Reload. 


\begin{lstlisting}[caption={Warping Path with Output Instruction}\label{lst:output},language=C] 
/*calculate the Euclidean distance of the 
three warping directions, choose the one 
with minimal distance cost*/
fp = fopen("warpingpath.txt","w");
double i1j = DistanceCost(i+1, j);
double i1j1 = DistanceCost(i+1, j+1);
double ij1 = DistanceCost(i, j+1);

if (i1j < i1j1 && i1j < ij1){
        prev_cost = i1j;
        fprintf(fp, "i+1\n");}
  
else if (ij1 < i1j && ij1 < i1j1){
        prev_cost = ij1;
        fprintf(fp, "j+1\n");}
    
else{
        prev_cost = i1j1;
        fprintf(fp, "i+1&j+1\n");}
\end{lstlisting}

\section{Vulnerability of Dynamic Time Warping-based ECG Classification}
\label{sec:vul}
This section will present the correlation between warping paths and labels of ECG traces. To achieve this, we consider the ECG datasets from the UC Riverside time series repository \cite{UCRArchive2018}, i.e., ECG200 \cite{olszewski2001generalized}. Following, we will first demonstrate the warping path distribution among $i+1$, $j+1$, $i+1\&j+1$ for ECG sequences from different classes. 

We explore whether warping paths of DTW are correlated to the labels of input ECG samples. To achieve this, we record the warping directions, i.e. $i+1$, $j+1$, $i+1\&j+1$ for samples with the same labels and samples with different labels. Taking ECG200 as an example, we will collect the warping path directions for DTW between two normal samples, two abnormal samples, one normal and one abnormal samples. We insert an output instruction in three warping direction codes to obtain clear warping paths as shown in List \ref{lst:output}. For simplicity reasons, we will name the three types of DTW inputs as: normal, abnormal, and hybrid. Normal means two ECG sequences DTW calculates are normal samples; abnormal means two ECG sequences DTW calculates are abnormal samples; hybrid means one ECG sequence is normal and the other is abnormal. We count the occurrences of each warping direction per 1000 warping choices. Since the length of ECG samples is 100, the total warping choices between two ECG sequences has around 100*100 times. 

\begin{figure}[!htb]
    \centering
    \includegraphics[width=0.47\textwidth]{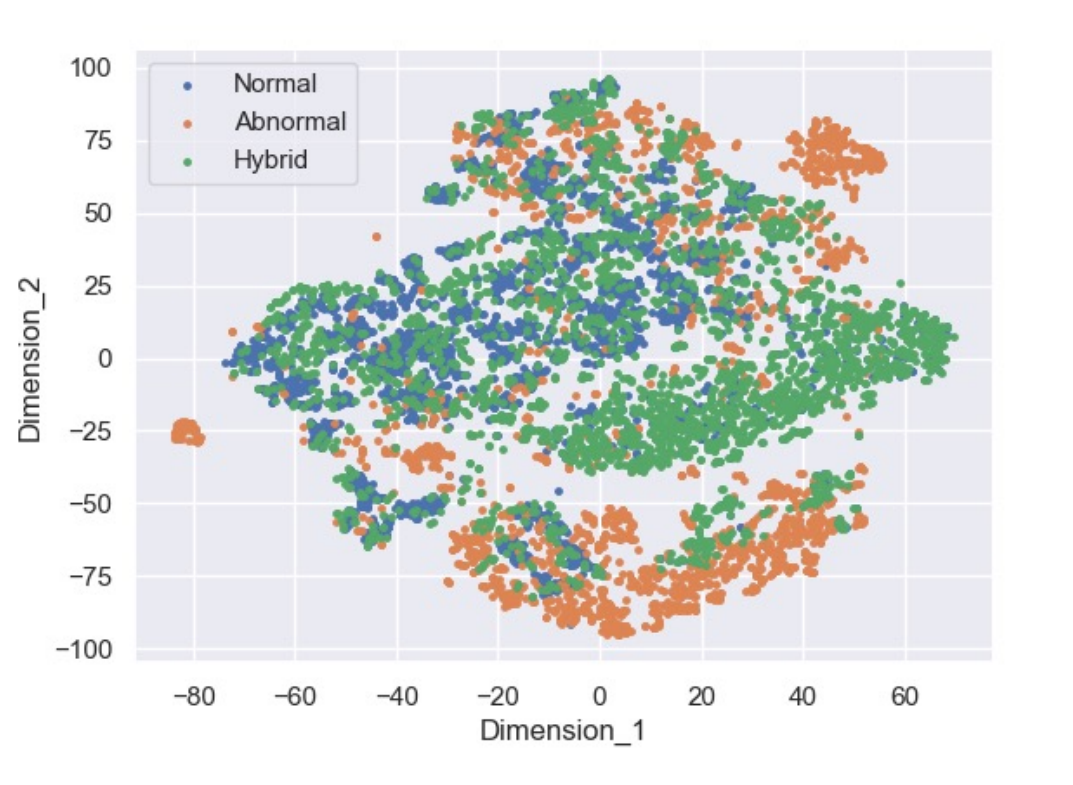}
    \caption{DTW warping selections t-SNE plot among normal, abnormal, and hybrid types of input ECG samples.}
    \label{fig:tnse}
\end{figure}

We evaluate the total number of the three path warping directions, i.e., $i+1$, $j+1$, $i+1\&j+1$, for three types of ECG input samples sent to DTW. We leverage t-SNE \cite{van2008visualizing} to explore how separable the warping directions are among DTW's two ECG input samples with different labels. t-SNE can help convert the warping selections into a two-dimension plot as shown in Figure \ref{fig:tnse}. We find that the warping selections are splittable among different ECG samples. This also aligns with the intuition that samples with the same labels are likely to be warped similarly. The divisibility of warping selections gives the opportunity to deduce the labels of the two ECG samples sent to DTW based on the observations of warping directions. The root cause behind the warping path distribution differences among the three types of inputs is the warping direction selection algorithm. As shown in List \ref{lst:output}, the distance cost decides the warping direction selections among $i+1$, $j+1$, $i+1\&j+1$. Hence, warping direction selection still contains the information on distance cost, which is also the basis of DTW-based classification.

\section{Attack Implementation}
Based on the vulnerability described in Section \ref{sec:vul}, we implement a cache-based side-channel attack Flush+Reload to spy on the warping path directions and infer the labels of ECG samples sent to DTW model. The implementation includes two main parts: a warping path collection and a machine learning-based classification model. 
\subsection{Threat Model}
\label{sec:threatmodel}
The victim is a DTW classification model for predicting ECG samples. The presented attacker's goal is to infer the labels of ECG samples sent to DTW by monitoring the warping selections. We leverage the cache-based Flush+Reload to spy the computations of DTW and assume that the attacker (without sudo access) resides in the same physical machine with the victim DTW model. Such co-location can be achieved in cloud by leveraging recent work \cite{fang2021repttack} or on edge devices by luring users to download malware. We assume that the attacker does not have direct access to input ECG samples, the DTW classification model, and the prediction results, meaning the ECG classification model is secure at software level. To build the correlation between side-channel observations and ECG samples' labels, the attacker is also supposed to have access to inquiry the victim DTW-based ECG classification model and have its own labeled ECG samples to inquire about the victim model. 

The threat model we considered is practical in the real world as well. When companies launch their ECG monitoring and classification products, the attacker can purchase one to insert Flush+Reload and inquiry it with its own labeled ECG data to collect warping paths for building the correlation between side-channel observations and ECG samples' labels. Once the attacker lures users to download malicious codes which contains Flush+Reload, the correlation will be exploited to deduce labels of other benign users' ECG samples.




\begin{table}[!htb]
    \centering
        \caption{RandomForest classification accuracy with different num of splits and warping directions.}
    \begin{tabular}{|c|c|c|c|c|}
    \hline
  Num of Splits&$i+1$&$j+1$&$i+1\&j+1$&Full Warping Path\\\hline
100& 81.1&83.7&84.8&85.1\\\hline
10&71.5&69.2&74.3&79.4\\\hline
1&37.4&43.6&38.5&53.1\\\hline
    \end{tabular}

    \label{tab:accwarp}
\end{table}


\subsection{Acquiring Path Selection via Flush+Reload}
We will collect the warping paths observed by Flush+Reload with the attacker's ECG samples and employ the dataset to train a classification model for deducing labels. Firstly, we implement Flush+Reload targeting the warping direction selection codes as shown in List \ref{lst:output}. However, one challenge is that the memory locations of the three direction selection codes are too close. Flush+Reload cannot obtain the execution observations for the three warping directions. Hence, we can only monitor one of the three path selections and sum the occurrences at every specific interval. The selected monitoring warping direction and interval of Flush+Reload traces can be determined based on the analysis of ECG warping paths. As shown in Table \ref{tab:accwarp}, we investigate the impact of choosing different warping directions and the number of splits on the labels' classification accuracy. The number of splits will then be used to divide one warping path trace into multiple parts, each of which has the same time interval. When the number of splits is 1, the trace is not divided. Table \ref{tab:accwarp} shows that the classification accuracy increases with the number of splits increasing from 1 to 100. Another observation is that the full warping path gives the highest accuracy for all three split settings. Nevertheless, selecting the optimal monitored warping direction can minimize the accuracy gap. In response, this work chooses 100 as the number of splits and $i+1\&j+1$ as the monitored warping direction since $i+1\&j+1$ outperforms $i+1$ and $j+1$ for differentiating ECG labels. 

\subsection{Machine Learning-based Classification Model}
We split the whole dataset collected in the prior step into 70\% and 30\% for training and testing sets. Additionally, we conduct ten-fold cross-validation on the training dataset to build a robust classifier with high accuracy in case of an over-fitting issue. Furthermore, we consider a vast range of ML classifiers in this work to select the optimal one with the highest accuracy. Five classifiers are selected: Tree-based RandomForest (RF), SVM-based SMO, rule-based JRiP, Multi-Layer Perceptron (MLP), and lazy learning-based IBK. We utilize Weka \cite{hall2009weka} to implement the five classification algorithms. 

\subsection{Online Attacking Procedure}
Once the classification model is trained, the attacker is ready for online deployment. The online attacking procedure includes the following steps:
\begin{itemize}
    \item Insert Flush+Reload malicious codes into target devices that also conduct DTW-based ECG classification services.
    \item Pre-process Flush+Reload observations and split each trace into 100 splits with equal time intervals.
    \item Send the processed data to the remote classifier owned by the attacker.
    \item Deduce the labels of DTW's input ECG samples with the classifier. The leaked labels can be used for other malicious activities.
\end{itemize}

\begin{table*}
    \centering
       \caption{Confusion Matrix with RF. }
    \begin{tabular}{|c|c|c|c|c|c|}
    \hline
         \backslashbox{Actual}{Predicted}&\makebox{Normal}&Abnormal&Hybrid &False Positive&Attacking Success Rate(\%) \\\hline
         Normal&2204&2&426&0.143&83.7\\\hline
         Abnormal&6&643&48&0.001&92.3\\\hline
         Hybrid&470&5&2166&0.142&82.0\\\hline
    \end{tabular}
    \label{tab:cmmlp}
\end{table*}
\section{Evaluation}
\subsection{Experiment Setup}
\subsubsection{Attack Platform}
All evaluations are done on a Dell desktop with 12  Intel core I5-10400 cores and a three-level cache system. It has isolated L1 and L2 caches with 192 KB and 1.5MB respectively. L3 cache memory is shared among all cores and has 12 MB capacity. The inclusiveness of L3 allows attackers to observe cache behaviors of victim applications running on the same machine. In our presented work, we leverage Flush+Reload, which exploits the shared L3 cache.

\subsubsection{Victim Dataset}
As discussed in Section \ref{sec:bg}, we select ECG 200 dataset from UC Riverside Time-series repository \cite{UCRArchive2018}. ECG 200 has 200 samples, with 133 labeled normal and 67 labeled abnormal, where the length of each sample is 100. Hence, ECG 200 gives us \(\frac{133*132}{2}\) paths for DTW computation of normal samples, \(\frac{67*66}{2}\) paths for DTW computation of abnormal samples, and \(133*67\) paths for DTW computation between hybrid samples. The attacking success rate, i.e., classification accuracy, is used to measure the effectiveness of the presented attack.

\begin{figure}[!htb]
    \centering
    \includegraphics[width=0.45\textwidth]{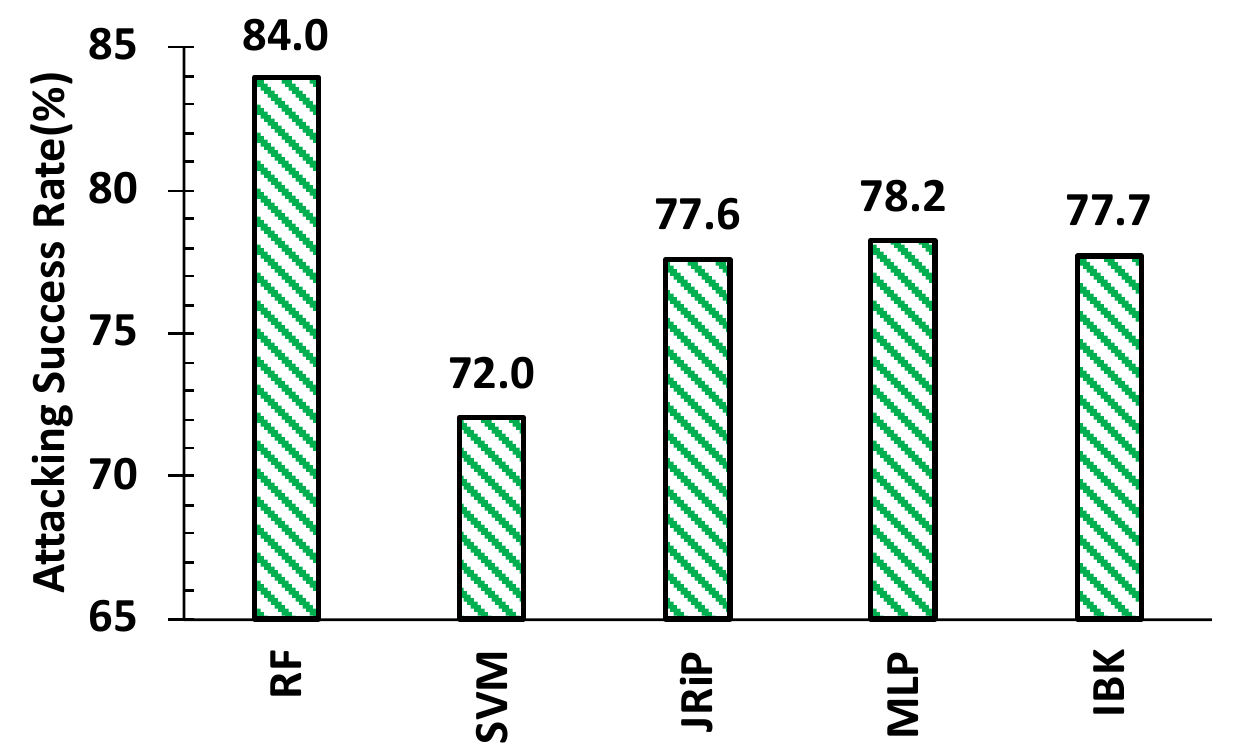}
    \caption{Attacking success rate with different classifier.}
    \label{fig:mlacc}
\end{figure}

\subsection{Attack Success Rate with Various Classifiers}
We explore different classification algorithms to select the optimal one for deducing victim models' inference results based on Flush+Reload observations. The attacking success rate of five classification algorithms from different categories is presented in Figure \ref{fig:mlacc}. We find that all five classifiers effectively infer the input ECG samples' labels, achieving above 70\% attacking success rate. Among them, RF outperforms the rest four and receives an 84.0\% success rate, indicating that the Flush+Reload observations of the warping path are highly related to the labels of ECG samples. Based on the results, RF is the most effective classifier for attackers to infer labels and will be used in the rest of the evaluation. 

What's more, the classification accuracy with side-channel observations for $i+1\&j+1$ directions is only slightly lower than the one with actual warping paths in Table \ref{tab:accwarp}, with 84.0\% and 84.8\% respectively. For the warping direction $i+1\&j+1$, a minor classification accuracy gap occurs because Flush+Reload observations still contain noises due to the transient executions or interrupts from other running programs, though Flush+Reload is designed to spy on specific behaviors in a program.

\subsection{Attacking Confidence}
Besides the overall attacking success rate, we also evaluate the confidence for the three types of DTW inputs with the confusion matrix of RF in Table \ref{tab:cmmlp}. We find that most misclassifications are from Normal and Hybrid. The inferences from the Abnormal class have the highest accuracy with 92.3\% and experience the lowest number of false positive rate with 0.001. Though the overall attacking success rate is 84.0\% with RF, the presented attacker can achieve higher accuracy of 92.3\% for abnormal class with a lower false positive rate of 0.001. The confusion matrix further highlights the leakage threats of DTW-based ECG classification via monitoring warping paths with side-channel attacks.

\section{Defense Discussion}
It is important to deploy effective defense approaches to address the presented side channel-assisted privacy attack on ML-based medical applications. 

One solution is to deploy effective detection modules and obtain the awareness of undergoing side-channel attacks. Prior works \cite{wang2020scarf,wang2020phased,wang2021enabling,wang2021evaluation,wangicmla,20230092190} have proposed leveraging hardware performance counters to build SCAs detectors and protect computer systems from privacy leakage. Another way to countermeasure the privacy compromise is designing mitigation methods and disrupt side-channel observations to secure medical information.
Several recent works have proposed to modify cache hierarchy or cache memory architecture to mitigate SCAs. Cache partitioning techniques \cite{page2005partitioned,domnitser2012non,liu2016catalyst} are proposed to mitigate cache-based side-channel attacks by statically or dynamically partitioning cache memory for each application process, thereby SCAs are not able to observe "side-channel information" of victim applications. Some works employ access randomization \cite{wang2007new,wang2008novel,liu2014random} which primarily randomizes cache interference, re-mapps cache indices, or replaces demand fetch with random cache fill to eliminate security vulnerabilities in 
the hardware architecture. The rest defenses approaches \cite{wang2020mitigating,wang2020hybrid,wang2022applied} proposed to adjust the settings of hardware and system parameters, including frequency and prefetchers, to increase the noise in side-channel observations.

\section{Conclusion}
ECG has been extensively used to measure the heartbeat and diagnose irregular heart rhythms. Machine learning-based classification techniques empower automatic abnormal detection and significantly reduce the demand for medical personnel. However, they also bring new security challenges. This work uses the time series classifier DTW as an example and presents the correlation between warping directions and ECG samples' labels. We leverage Flush+Reload to monitor the warping directions, prototype an attack that exploits the correlation and employ machine learning classifiers to deduce the labels of DTW's inputs. We find the presented attack can achieve up to 84.0\% attacking success rate. In the confusion matrix of the proposed attack, we further find that the attacking success rate for abnormal inputs is the highest, having 92.3\%. This work highlights the privacy leakage of DTW-based ECG classifications and sheds light on future defense approaches.
\bibliographystyle{IEEEtranS}
\bibliography{main.bbl}

\end{document}